\documentstyle[12pt,aasms4]{article}

\def\ltsima{$\;\buildrel < \over \sim \;$}
\def\simlt{\lower.5ex \hbox{\ltsima}}
\def\gtsima{$\;\buildrel > \over \sim \;$}
\def\simgt{\lower.5ex \hbox{\gtsima}}
\begin{document}
\title{Photon Bunching at TeV Energies}

\author{M. Harwit\altaffilmark{1}} 

\altaffiltext{1}{511 H Street SW, Washington, DC 20024-2725; also Cornell
University.}

\begin{abstract}

Harwit, Protheroe, and Biermann (1999) recently proposed that Bose-Einstein photon bunching 
might significantly affect the interpretation of Cerenkov counts of TeV gamma photons.  Here,
we show that a combination of two recent results of Aharonian et al. (2000) and Aharonian et al.
(2001) permits us to set new, more stringent upper limits of $\lesssim 10\%$ on the fractional
amount of  photon bunching in the 7-10 TeV radiation from Markarian 501.  Potential bunching
at even higher energies should nevertheless continue to be investigated for this and other TeV
sources, since a clear understanding of TeV energy spectra is required to unambiguously
determine the spectral energy density of the mid-infrared extragalactic background. 
\end{abstract}


\section{Introduction}

Harwit, Protheroe and Biermann (1999, hereafter Paper I) have proposed that some of the highest
energy TeV Cerenkov gamma-showers attributed to single photons arriving from distant blazars
might actually be produced by bunched photons at lower energies.  Lower energy photons can
penetrate greater distances through extragalactic space without suffering annihilation through
collision with photons from the intergalactic infrared background and producing
electron-positron pairs. The bunching proposed in Paper I was just the ordinary Bose-Einstein
bunching well known at lower energies.  Levinson (2000, 2001) has recently re-analyzed this
possibility, basing his study largely on a laser mechanism which, however, obeys very different
photon statistics. This letter provides a more detailed analysis of photon statistics, and goes
on to discuss the specific properties of the TeV radiation from Mrk 501. Section 2 of this letter
provides a quantitative treatment of photon fluctuations and correlations from incoherent sources. 
Section 3 sets new upper limits on the Bose-Einstein bunching of TeV photons from Mrk 501.
Section 4 touches on photon fluctuations in coherent processes like the laser mechanisms
investigated by Levinson (2000, 2001), and shows how these differ with specific reference to
TeV gamma photons. Section 5 stresses the importance of more stringent TeV observations in
order to better define the infrared extragalactic background.  A final section summarizes the main
conclusions.

\section{Quantitative Treatment}

The photon fluctuations in a beam of radiation can be derived in a number of different ways
(Hanbury Brown \& Twiss, 1957, 1958a, 1958b, 1958c; Purcell, 1956; Wolf, 1957; Mandel,
1958; Harwit, 1960).  Let us assume first that the radiation is blackbody at some temperature T. 
Then, the Einstein-Fowler equation tells us that 
\begin{equation}
\langle(\Delta E)^2\rangle = kT^2\frac{\delta \langle E \rangle}{\delta T} = h\nu\langle E\rangle
\biggl [ 1 + \frac{1}{exp(h\nu /kT) - 1} \biggr ]
\end{equation}
where the brackets $\langle \rangle$ denote ensemble averages, $T$ is the source temperature,
$E$ is
the energy of the individual photons and $\nu$ is their spectral frequency.  From this we can
immediately obtain the fluctuation $\langle (\Delta N)^2\rangle$in photon number $N$  in a
given volume and its relationship to the mean value $\langle N\rangle$
\begin{equation}
\langle(\Delta N)^2\rangle = \langle N\rangle\biggl [1 + \frac{1}{exp(h\nu /kT) - 1} \biggr ] = 
\langle N\rangle\biggl [1 +\langle N\rangle / g \biggr ]
\end{equation}
where g is the number of phase cells in the volume occupied by the photons.  
\begin{equation}
g = 2A\tau c \Omega \nu_0^2 \Delta \nu c^{-3}
\end{equation}
and $A$ is the aperture of the telescope, $\tau$ is the time interval over which a single
measurement in the ensemble of measurements of $N$ is made, $\Omega$ is the solid angle
subtended by the source, $\nu_0$ is the central frequency and $\Delta \nu$ the radiation
bandwidth. The numerical factor 2 accounts for the two possible states of polarization.

If the radiation incident on the light-gathering telescope is coherent, in the sense that the radiating
source just fills the telescope's diffraction angle at frequency $\nu_0$, the transverse dimension
of the incoming beam is simply the cross sectional area of a single phase cell and, for fully
polarized
photons, the number of phase cells along the direction of the beam is  $g \sim  \tau\Delta\nu$.
Though we have derived this relationship for blackbody radiation, Mandel, Sudarshan and Wolf
(1964) showed that it applies to any quasi-monochromatic coherent beam of radiation, regardless
of whether or not it originates in a statistically stationary thermal source, as long as the wave
amplitudes in the beam are represented by Gaussian random variables (Mandel, 1963). 

Equations (1) and (2) specifically, do not apply to lasers or masers operating well above their
thresholds; this is further discussed in section 4, below.

In practice, we do not directly measure numbers of photons, but rather instrumental counts
produced by the photon flux.  If the counting efficiency of the instrumentation is $\epsilon$, the
fluctuation in the number of photoelectrons, $N_{\epsilon}$, becomes 
\begin{equation}
\langle(\Delta N_{\epsilon})^2\rangle = \epsilon \langle N\rangle\biggl [1 +\epsilon \langle
N\rangle / g
\biggr ]
\end{equation}
where the values of $\epsilon$ and $g$ characterizing the instrumentation and the incoming
stream of radiation may be experimentally determined. This permits us to calculate the true
fluctuations in the incoming radiation beam $\langle(\Delta N)^2\rangle $ and their relation to the
mean number of photons $\langle N\rangle$ observed in a given time interval. For radiation with
a degree of polarization $0\leq P\leq 1$ the corresponding expression is (Wolf, 1960)
\begin{equation}
\langle(\Delta N_{\epsilon})^2\rangle = \epsilon \langle N\rangle\biggl [1 +\epsilon (1+P^2)
\langle N\rangle / g \biggr ]
\end{equation}
While fluctuations can, in principle, be measured in single beams, a more reliable determination
is often obtained by dividing the incoming radiation so half falls on one detector and half
on another, for instance through use of a beam splitter, and determining the correlations or
coincidence counts in the radiation incident on the two detectors.  The correlation in the counts
on the detectors, respectively labeled by subscripts 1 and 2, then becomes
\begin{equation}
\langle(\Delta N_{\epsilon 1}\Delta N_{\epsilon 2})\rangle = \epsilon^2\frac{1}{2}(1+P^2)
\langle
N_1\rangle\langle N_2\rangle / g
\end{equation}
Hanbury Brown et al. (1967) showed that for sources small compared to a telescope's angular
resolving power, the light could be gathered by two telescopes separated by variable distances.
When the separation $D$ between telescopes became comparable to the diameter of a telescope
that, at spectral frequency $\nu$ would just be able to resolve a star, the correlations in counts
would begin to drop, giving the star's angular diameter as $\theta \sim \lambda/D$, where
$\lambda = c/\nu$ is the wavelength and $c$ is the speed of light. While Hanbury Brown et al.
were working with optical telescopes, following earlier observations in the radio domain, the
same photon bunching could also be observable with Cerenkov telescopes operating at the
highest gamma frequencies.

To quantitatively see how the fluctuations are related to the coincident arrivals posited in Paper I
we need to look at the probability $P(N)$ of finding $N$ photons per phase cell as a function of
mean photon number in the cell.  Mandel (1963) has shown this to be
\begin{equation}
P(N) = \frac{1}{(1+\langle N\rangle )(1+\frac {1}{\langle N\rangle})^N}
\end{equation}
This is the Bose-Einstein probability distribution.

Mandel and Wolf (1995) explicitly show this expression to hold for response times $\tau$ short
compared to the correlation time of the intensity of the radiation.  This is the case for a Cerenkov
array detecting individual TeV photons arriving at intervals long compared to the time resolution
 $\tau$ of the array.  The meaning attached to $\Delta \nu$ may then be thought of as the
uncertainty in the frequency of each counted photon, given by $\Delta\nu \sim \tau ^{-1}$ rather
than, as assumed by Levinson (2001), the far larger bandwidth of photons emitted by the source,
which becomes apparent only when an entire ensemble of Cerenkov events is analyzed.  

For an aperture just large enough to resolve the solid angle subtended by the source, and for the
fast response time of the Cerenkov array and the sparse flux of arriving TeV photons, $\tau
\Delta\nu = 1$, and $g$ also takes on the value 1.

Paper I has suggested that Cerenkov radiation observed at the highest TeV energies might
arise from two or more photons arriving in coincidence from an extremely energetic source. It
also showed how this possibility could be tested through detailed observations of the
development of the Cerenkov shower in the atmosphere.   

\section{The Crab Pulsar and Markarian 501}

Two crucial papers now need to be cited: Aharonian et al (2000) and Aharonian et al (2001).

The first of these papers reports informative tests along the lines proposed in Paper I.  The
authors find that for the Crab pulsar and the blazar Mrk 501 coincident arrival of photons within
the angular resolution of the Cerenkov array, i.e. within a requisite width transverse to the
direction of arrival at the altitude of initial impact, can be ruled out at roughly a 1 $\sigma$ level
for the highest energy $\sim 15\,$TeV photons.  The authors provide an upper limit on the
bunching based on the anticipated development of the Cerenkov shower.  At a 90\% confidence
level they set an upper limit of 65\% on the number of photon pairs, N = 2,  that could be added
to the shower without significantly changing the shower structure from that produced by single
photons having an equal combined energy.  At a similar level of confidence the authors rule out
that more than 30\% of the showers produced at any given energy could be produced by bunches
of N = 10 photons, each at an energy one tenth of that of their combined energies.  

When Paper I was written, one could not rule out that the apparent TeV spectrum of a source was
strongly influenced by photon bunching.  The apparent spectral energy distribution at the highest
energy could, in that case,  have been entirely due to highly bunched photons masquerading as
individual photons with much higher energies.  As a result of the work of Aharonian et al.
(2000), we now know that this possibility can be largely ruled out for Mrk 501, and that the
observed Cerenkov spectrum well represents the actual spectral energy distribution of arriving
photons in the $\sim1$ to 10 TeV range.  However, the concern about bunching is greatest at the
highest TeV energies, where traversal across extragalactic space is most severely affected by
annihilation through impact against intergalactic infrared photons. 

The second paper of importance is that of Aharonian et al. (2001), which provides an improved
set of spectra for Mrk 501 in both the high and the low state.  Knowledge of the spectral shape
permits us to derive an even more stringent upper limit on the photon bunching than given by
Aharonian et al. (2000).  This can be seen by considering  the steep decline in the spectrum of
Mrk 501 at the highest energies $\gtrsim 10\,$TeV.  For the year 1997, the observed drop in the
number of photons per TeV interval was approximately a factor of $\sim 20$ in going from
energies of 8\,TeV  to 16\,TeV.  If as few as 2 out of every 20 photons at 8\,TeV arrived in
coincidence, all of the 16\,TeV photons could be accounted for.  Since there are so many lower
energy photons, the fraction $f$ of these that needs to be bunched in order masquerade as single
higher-energy photons could be quite small.  A fraction $f\sim 0.1$ appears quite compatible
with the results of Aharonian et al (2000).  From equation (3) we then have,  $\langle N\rangle /g
\gtrsim f \sim 0.1$. 
                                                  
For $\langle N\rangle = 0.1$ the probability given by equation (7) for the arrival of an empty
phase cell is 0.909, for the arrival of one photon it is 0.083, for 2 photons it is $\sim 0.0075$, and
for each additional photon a further factor of $\sim 11$ less. 

A point noted by Levinson (2001) should still be mentioned.  For a steady incident flux,
Levinson set  $\langle N \rangle /(A\tau)$ equal to the flux, or the number of Cernekov counts
per unit area per unit time, and argued that for the observed TeV flux $F \sim
10^{-10}\,$cm$^{-2}$\,s$^{-1}$
this corresponds to $\langle N\rangle \lesssim 10^{-12}$ for a ``spot size"  20\,m by 20\,m $\sim
4\times 10^6\,$cm$^2$ and $\tau \sim 10^{-9}$\,s.  

This would be correct if the incident flux were steady.  However, a steady flux is not expected. 
For an intermittent flux with low duty cycle, $\langle N\rangle$ can be many orders of magnitude
higher.   

A low duty cycle is expected because the conditions that permit the production of
gamma rays at the very highest energies are very special.  Most models attempting to account for
the TeV radiation assume inverse Compton scattering off highly relativistic electrons.  Paper I,
for example, assumed a Lorentz factor $\Gamma$ for a relativistic jet and  a Lorentz
factor $\delta$ for the relativistic electrons comoving with the jet.  To produce gammas at the
very highest energies $E$, a product $(\Gamma \delta)^2\sim 10^{16}(E/(TeV)$, and a
fortuitous alignment of the two directions of motion was required.  The inverse Compton scatter
would then beam these photons into solid angles of order $(\delta\gamma)^{-2}\sim 10^{-16}$
sr.  The required alignment of the two sets of relativistic velocities, and the narrow beaming
angles in what are likely to be turbulent jets,  strongly suggest that conditions favorable to
producing the highest energy TeV photons are short-lived and have extremely low duty cycles. 
The arrival of TeV photons will be episodic, and $\langle N\rangle$ should not be derived from a
mean flux.  A value of  $\langle N\rangle\sim 0.1$ accordingly seems well within expectations.

Even though bunching may occur both at lower and higher energies, it is
more readily observed at higher energies because higher energy photons from a given source are
laterally more tightly confined within a single phase cell.  If the apparent aperture, $D$, of the
Cerenkov telescope ---  i.e. the limiting ``spot size" in the upper atmosphere that can be  resolved
by the Cerenkov array --- is sufficiently large to just resolve a given source at some energy $E$,
then two photons at energy $E$ belonging to the same phase cell can both give rise to the same
shower.  In contrast, as explained in connection with equation (6) above, a pair of photons, each
having half this energy, namely $E/2$, will be distributed over an area of diameter $2D$, 4 times
as large as the ``spot size" and the probability of both photons contributing to the Cerenkov pulse
is reduced by a factor of 4.

In summary, the upper limits on photon bunching that Ahraronian et al. (2000) provide, tell us
that the observed photon number distribution $N(E)$ is the actual spectral distribution, and that
bunched photons masquerading as single higher-energy photons can only affect the spectral
energy distribution to a minor extent.  Given this, the rapidly declining photon number at
high energies can be used to further constrain the number of coincidentally arriving photons,
simply because a small fraction of bunched photons at energies $E/2$ would suffice to provide
all of the photons arriving with energies $E$

Further observations, nevertheless, should be undertaken to rule out even this low occurrence of
bunching at the highest energies because it is just these photons that are most critically
affected by annihilation by mid-infrared photons at $\sim10 - 100\,\mu$m, for which direct
observations of the extragalactic infrared background are particularly difficult.

\section{Fluctuations from Cosmic Lasers or Masers}

Levinson (2000, 2001) has investigated whether coincident photons could reasonably be
expected to be produced by plane parallel lasers at TeV energies.  The two papers by Aharonian et al. (2000, 2001) taken together now show that the observed TeV spectrum cannot be due
to a superposition of large numbers of monochromatic photons. Accordingly, the flux generated
at the source must be largely broad-band.  This makes a laser model highly unlikely.

Nevertheless, the photon statistics of laser and incoherent radiation should be examined in order
to better understand one point emphasized  in Levinson's two papers.  As Levinson points out,
the photons from a laser with diameter $D$ must diverge with an angle $\sim \lambda /D$
because of diffraction, and the arriving beam at Earth can never be smaller than $D + \lambda
L/D$, where $L$ is the distance to the source. For Mrk 501 at a distance of $\sim$ 130 Mpc
Levinson concludes that at TeV energies ``the minimum spot size of a beam of photons arriving
at Earth is of the order
of a few kilometers." This is much larger than the ``spot size" roughly 20 m in diameter within
which Cerenkov measurements would be able to register photons arriving simultaneously.

As already mentioned, the equations presented in section 2 specifically do not apply to plane
parallel lasers operating well above threshold because their emission is not represented by a
Gaussian random process, but is coherent.   Under these conditions, the Bose-Einstein
fluctuations, represented by the second term on the right in equations (5) and (6) are not
expected.  Only the first term on the right is found and takes the form of a classical Poisson
distribution.  Early experiments by Smith and Armstrong (1966) were able to verify that for
lasers operating well above threshold the Bose-Einstein fluctuations, and with them the
coincidentally arriving photons, disappear. 

For lasers the probability for a coincident pair of  photons, as contrasted to a single photon,
falling within the limited area viewed by existing Cerenkov arrays is then proportional to the
ratio of the area of the array to the area of the laser ``spot size" --- roughly a factor of $\sim
10^{-4}$, far lower than the fraction expected for bunched photons arriving from an
incoherent source of identical dimensions.

The reasons for this difference are straightforward.  

The output of a plane parallel laser originates in a single spontaneous
transition in resonance with an enclosing Fabry-Perot cavity. This seed
radiation is amplified in a cascade of stimulated emissions that produce a
beam of photons all of which are quantum mechanically identical.  What emerges
from the laser is a coherent, plane parallel beam effectively produced in a
point source at infinity, and constricted to pass through the aperture of the
laser end window.  Such an ensemble interferes with itself to diverge in a
diffraction limited beam that becomes wider the smaller the laser exit
aperture.

In contrast, for a source that randomly emits photons, say through blackbody
emission, just the opposite is true.  The larger the original distant source,
the smaller the spatial region in which arriving photons with identical
quantum properties are confined.  This is because the angles of arrival, i.e.
the transverse momenta of the photons are quite uncertain (as they would also be for a laser beam 
focused by a lens). This permits high spatial localization limited only by the Heisenberg
uncertainty principle.  Processes of the kind advocated in Paper I, in which light is incoherently
inverse-Compton scattered over a sizeable source area, similarly satisfy the conditions for which
equations (1) to (7) apply.

\section{Discussion}

As stated in Paper I, it is difficult to know what types of astrophysical processes could produce
bunched TeV photons in blazars, and one may argue about the plausibility of the specific
process outlined in Paper I.  Nevertheless, the very highest energy TeV gamma events should
continue to be more precisely analyzed to determine whether a significant fraction is produced by
bunched photons. Even if coincident arrival cannot account for most of the 5-10 TeV photons
from Mrk 501, it will be important to place more stringent upper limits on photons in the 10 - 20
TeV range, where the photon flux dramatically falls off (Aharonian et al., 2001). If only a minor
fraction of photons at energies of order 7 to 8 TeV arrives in pairs, so that $\langle N\rangle /g$
in equations (5) or (7) is merely of order 0.1, this could already be sufficient to account for a
significant fraction of the observed 15 TeV photons.

Because direct observations of the extragalactic infrared background are made difficult by the
presence of zodiacal glow, scattered starlight, and high-latitude Galactic cirrus emission, TeV
observations may in the long run provide the most precise measure of the infrared background
and its spectral energy distribution (Stecker, 2001; Dwek \& de Jager, 2001).  For this to happen,
however, we will have to be sure that the spectrum of blazars at different distances is well
understood, and that Cerenkov measures do not slant these spectra toward higher energies by
mistaking photon pairs of slightly lower energies for single photons at the very highest energies. 

\section{Conclusion}

We have shown that the two papers by Aharonian et al (2000, 2001) taken together permit us to
restrict the number of coincidentally arriving TeV photons more precisely than these authors'
first paper.  For Mrk 501 the number of paired photons arriving at some of the highest energies
$E\sim 7 - 10\,$TeV cannot exceed $\sim 10\%$ without exceeding the total flux observed at
energies $2E \sim 14 - 20\,$TeV, where the 1 $\sigma$ error bars are still quite large. Continued
observations could, however, rule out even this eventuality. This is essential because the accurate
determination of the mid-infrared extragalactic background radiation critically depends on a
proper understanding of the actual energy spectrum of the highest-energy TeV photons.  While
we do not currently know of any astrophysical process certain to produce bunched photons at
TeV energies, the possibility can only be observationally ruled out. 

\section{Acknowledgments}

I would like to thank Prof. Emil Wolf for guiding me to a number of important articles I had not
previously seen, and for commenting on an early draft of this paper.  I am also pleased to
acknowledge an exchange of ideas, during the preparation of this letter, with Raymond 
Protheroe and Peter Biermann, my co-authors on Paper I.  My research in infrared astronomy is
supported by NASA grants and contracts.

\section{References}

Aharonian, F. et al. 2000, ApJ 543, L 39

Aharonian, F. et al. 2001, ApJ 546, 898

Dwek, E. \& de Jager, O. C. 2001, IAU Symposium 204, {\it The Extragalactic Background and
its Cosmological Implications} Martin Harwit, and Michael G. Hauser, eds., Astronomical
Society of the Pacific, p 398

Hanbury Brown, R. \& Twiss, R. Q. 1957 Proc. Roy. Soc. London, A, 242, 300

Hanbury Brown, R. \& Twiss, R. Q. 1958a Proc. Roy. Soc. London, A, 243, 291

Hanbury Brown, R. \& Twiss, R. Q. 1958b Proc. Roy. Soc. London, A, 248, 199 

Hanbury Brown, R. \& Twiss, R. Q. 1958c Proc. Roy. Soc. London, A, 248, 22 

Hanbury Brown, R., David, J., Allen, L. R., \& Rome, J.M. 1967, MNRAS 137, 393

Harwit, M. 1960, Phys. Rev.120, 1551

Harwit, M., Protheroe, R. J., \& Biermann, P. L. 1999, ApJ 524, L91

Levinson, A., 2000 astro/ph 0007109

Levinson, A. 2001, ApJ 549, L67

Mandel, L. 1958, Proc. Phys. Soc. (London) 72, 1037

Mandel, L. 1963, Progress in Optics, II, 181

Mandel, L. \& Wolf, E. 1995, {\it Optical Coherence and Quantum Optics}, Cambridge
University Press, p 451

Mandel, L., Sudarshan, E. C. G., \& Wolf, E. 1964, Proc. Phys. Soc. 84, 435

Purcell, E. M. 1956, Nature 178, 1449 

Smith, A. W., \& Armstrong, J. A. 1966, Phys. Rev. Lett. 16, 1169

Stecker, F. W. 2001, IAU Symposium 204, {\it The Extragalactic Background and its
Cosmological Implications} Martin Harwit, and Michael G. Hauser, eds., Astronomical Society
of the Pacific, p 135

Wolf, E. 1957, Phil. Mag 2, 351

Wolf, E. 1960, Proc. Phys. Soc. 76, 424

\end{document}